\documentclass[12pt]{article}

\usepackage{url, graphicx, color, varioref, amsfonts, longtable, float,epsfig}

\newtheorem{theorem}{Theorem}[section]

\newenvironment{proof}[1][Proof]{\begin{trivlist}
\item[\hskip \labelsep {\bfseries #1}]}{\end{trivlist}}

\newcommand{\qed}{\nobreak \ifvmode \relax \else
      \ifdim\lastskip<1.5em \hskip-\lastskip
      \hskip1.5em plus0em minus0.5em \fi \nobreak
      \vrule height0.75em width0.5em depth0.25em\fi}

\title{Quantum Search Algorithm with more Reliable Behaviour using Partial Diffusion}

\author{Ahmed Younes\footnote {Birmingham, Edgbaston, B15 2TT, United Kingdom , axy@cs.bham.ac.uk} \,\,\,\,\,\,\, Jon Rowe \footnote {Birmingham, Edgbaston, B15 2TT, United Kingdom , jer@cs.bham.ac.uk} \\ School of Computer Science \\ University of Birmingham \\
\and
Julian Miller \footnote {York, Heslington, YO10 5DD, United Kingdom, jfm@ohm.york.ac.uk}\\  Department of Electronics\\   University of York\\  
}

\begin{document}
\maketitle
\begin{abstract}
In this paper, we will use a quantum operator which performs the inversion about the mean operation only 
on a subspace of the system ({\it Partial Diffusion Operator}) to propose a quantum search algorithm 
runs in $O(\sqrt{N/M})$ for searching unstructured list 
of size $N$ with $M$ matches such that, $1\le M \le N$. We will show that the performance of the algorithm is more reliable 
than known quantum search algorithms especially for multiple matches within the search space. A performance comparison with 
Grover's algorithm will be provided.

\end{abstract}


\section{Introduction}

Quantum computers \cite{Deutsch85,Feynman86,Lloyd93} are probabilistic devices, which promise to do 
some types of computation more powerfully than classical computers \cite{Bern93,simon94}. 
Many quantum algorithms have been presented recently, for example, Shor \cite{shor97} presented a quantum 
algorithm for factorising a composite integer into its prime factors in polynomial time. 
Grover \cite{grover96} presented an algorithm for searching unstructured list of $N$ items with quadratic 
speed-up over algorithms run on classical computers.

Grover's algorithm inspired many researchers, including this work, to try to analyze and/or generalize 
his algorithm \cite{Accardi00,boyer96,Bras00,Galindo00,Jozsa99,younes03}. Grover's algorithm perfomance is near to optimum for a 
single match within the search space, although the number of iterations required by the algorithm increases; 
i.e. the problem becomes harder, as the number of matches exceeds half the number of items in the search space \cite{niel00}  
which is undesired behaviour for a search algorithm since the problem is expected to be easier for multiple matches. 

In this paper, using a {\it partial diffusion operation}, we will show a quantum algorithm, which can find 
a match among multiple matches within the search space after one iteration with probability at least 
90\% if the number of matches is more than one-third of the search space. For fewer matches the 
algorithm runs in quadratic speed up similar to Grover's algorithm with more reliable behaviour, 
as we will see.

The plan of the paper is as follows:  Section 2 gives a short introduction to quantum computers. 
Section 3 introduces the search problem and Grover's 
algorithm performance. Section 4 and 5 introduce the proposed algorithm with analysis on its 
performance and behaviour. And we will end up with a conclusion in section 6.

\section{Quantum Computers}

\subsection{Quantum Bits}

In classical computers, a bit is considered as the basic unit for 
information processing; a bit can carry one value at a time (either 0 or 1). 
In quantum computers, the analogue of the bit is the quantum bit ({\it qubit} 
\cite{sch95}), which has two possible states encoded as $\left| 0 \right\rangle $ 
and $\left| 1 \right\rangle $; where the notation $\left| \,\,\, \right\rangle $ is 
called {\it Dirac Notation} and is considered as the standard notation of states in quantum 
mechanics \cite{dirac47}. For quantum computing purposes, the states $\left| 0 
\right\rangle $ and $\left| 1 \right\rangle$ can be considered as the 
classical bit values 0 and 1 respectively. An important difference between a 
classical bit and a qubit is that the qubit can exist in a linear 
superposition of both states ($\left| 0 \right\rangle $ and $\left| 1 
\right\rangle )$ at the same time and this gives the hope that quantum computers 
can do computation simultaneously ({\it Quantum Parallelism}). If we consider a quantum register with 
$n$ qubits all in superposition, then any operation applied on this register will be applied on 
the $2^{n}$ states representing the superposition simultaneously.

\subsection{Quantum Measurements}

To read information from a quantum register (quantum system), we must apply 
a measurement on that register which will result in a projection of the 
states of the system to a subspace of the state space compatible with the 
values being measured. For example, 
consider a two-qubit system $\left| \phi \right\rangle $ defined as follows:

\begin{equation}
\label{ENheq2}
\left| \phi \right\rangle = \alpha \left| {00} \right\rangle + \beta \left| 
{01} \right\rangle + \gamma \left| {10} \right\rangle + \delta \left| {11} 
\right\rangle ,
\end{equation}

\noindent
where $\alpha $, $\beta $, $\gamma $, and $\delta $ are complex numbers called the 
amplitudes of the system and satisfy $\left| \alpha \right|^2 + \left| \beta \right|^2 + \left| \gamma 
\right|^2 + \left| \delta \right|^2 = 1$. The probability that the first qubit of 
$\left| \phi \right\rangle $ to be $\left| 0 \right\rangle $ is equal to 
$\left( {\left| \alpha \right|^2 + \left| \beta \right|^2} \right)$. 
If for some reasons we need to have the value $\left| 0 \right\rangle $ 
in the first qubit after any measurement, we must try some how 
to increase its probability before applying the measurement. Note that, the new 
state after applying measurement must be re-normalized so the total probability is 
still 1.

\subsection{Quantum Gates}

In general, quantum algorithms can be understood as follows: Apply a series 
of transformations (gates) then apply the measurement to get the desired 
result with high probability. According to the laws of quantum mechanics and to keep the 
reversibility condition required in quantum computation, the evolution of 
the state of the quantum system $\left| \psi \right\rangle $ of size $n$ by 
time $t$ is described by a matrix $U$ of dimension $2^n\times 
2^n$ \cite{niel00}:

\begin{equation}
\label{ENheq4}
\left| {\psi '} \right\rangle = U\left| \psi \right\rangle ,
\end{equation}

\noindent
where $U$ satisfies the unitary condition: $U^{\dag} U = I$, where $U^{\dag}$ denotes 
the complex conjugate transpose of $U$ and $I$ is the identity matrix. For example, the $X$ gate ($NOT$ gate) is 
a single qubit gate (single input/output) similar in its effect to the classical $NOT$ gate. It inverts the 
state $\left| 0 \right\rangle $ to the state $\left| 1 \right\rangle$ and 
visa versa. It's $2\times2$ unitary matrix takes this form,

\begin{equation}
\label{ENheq5}
X = \left[ {{\begin{array}{*{20}c}
 0 \hfill & 1 \hfill \\
 1 \hfill & 0 \hfill \\
\end{array} }} \right],
\end{equation}

\noindent
and its circuit takes the form shown in Fig.(\ref{ENhfig1}). Notice that, from now on we assume that 
a horizontal line used in any quantum circuit represents a qubit and the flow of the circuit logic 
is from left to right. For circuits with multiple qubits, qubits will be arranged according to 
the notation used in the figure.

\begin{center}
\begin{figure}[H]
\begin{center}
\setlength{\unitlength}{3947sp}%
\begingroup\makeatletter\ifx\SetFigFont\undefined%
\gdef\SetFigFont#1#2#3#4#5{%
  \reset@font\fontsize{#1}{#2pt}%
  \fontfamily{#3}\fontseries{#4}\fontshape{#5}%
  \selectfont}%
\fi\endgroup%
\begin{picture}(1875,399)(4201,-2998)
\thinlines
{\color[rgb]{0,0,0}\put(4701,-2986){\framebox(600,375){}}
}%
{\color[rgb]{0,0,0}\put(5301,-2761){\line( 1, 0){300}}
}%
{\color[rgb]{0,0,0}\put(4701,-2761){\line(-1, 0){300}}
}%
\put(4926,-2836){$X$}%
\put(3101,-2836){$\left( {\alpha \left| 0 \right\rangle  + \beta \left| 1 \right\rangle } \right)$}%
\put(5676,-2836){$\left( {\beta \left| 0 \right\rangle  + \alpha \left| 1 \right\rangle } \right)$}%
\end{picture}
\end{center}
\caption{$NOT$ gate quantum circuit.}
\label{ENhfig1}
\end{figure}
\end{center}

Another important example is the Hadamard gate ($H$ gate) which has no classical equivalent; it produces 
a completely random output with equal probabilities of the output to be 
$\left| 0 \right\rangle $ or $\left| 1 \right\rangle $ on any measurements. 
It's $2\times2$ unitary matrix takes this form,

\begin{equation}
\label{ENheq6}
H = \frac{1}{\sqrt 2 }\left[ {{
\begin{array}{*{20}c}
 1 \hfill & \,\,\,\,1 \hfill \\
 1 \hfill & { - 1} \hfill \\
\end{array}
}} \right],
\end{equation}

\noindent
and its circuit takes the form shown in Fig.(\ref{ENhfig2}).

\begin{center}
\begin{figure}[H]
\begin{center}
\setlength{\unitlength}{3947sp}%
\begingroup\makeatletter\ifx\SetFigFont\undefined%
\gdef\SetFigFont#1#2#3#4#5{%
  \reset@font\fontsize{#1}{#2pt}%
  \fontfamily{#3}\fontseries{#4}\fontshape{#5}%
  \selectfont}%
\fi\endgroup%
\begin{picture}(1875,399)(4201,-2998)
\thinlines
{\color[rgb]{0,0,0}\put(4301,-2986){\framebox(600,375){}}
}%
{\color[rgb]{0,0,0}\put(4901,-2761){\line( 1, 0){300}}
}%
{\color[rgb]{0,0,0}\put(4301,-2761){\line(-1, 0){300}}
}%
\put(4526,-2836){$H$}%
\put(3750,-2836){$\left| x \right\rangle$}%
\put(5276,-2836){$\frac{1}{{\sqrt 2 }}\left( {\left| 0 \right\rangle  + \left( { - 1} \right)^x \left| 1 \right\rangle } \right)$}%
\end{picture}
\end{center}
\caption{Hadamard gate quantum circuit, where $x$ is any Boolean variable.}
\label{ENhfig2}
\end{figure}
\end{center}

Controlled operations are considered as the heart of quantum computing \cite{bare95}, Controlled-$U$ gate is the 
general case for any controlled gate with one or more control qubit(s) as shown in Fig.(\ref{ENhfig3_4_X}.a). It works as 
follows: If any of the control qubits $\left| {c_i } \right\rangle $'s ($1 \le i \le n - 1)$ is 
set to 0, then the quantum gate $U$ will not be applied on target qubit $\left| t \right\rangle $; i.e. 
$U$ is applied on $\left| t \right\rangle$ if and only if all $\left| {c_i } \right\rangle $'s are set to 1. 
The states of the qubits after applying the gate will be transformed according to the following rule:

\begin{equation}
\label{ENheq7}
\begin{array}{l}
 \left| {c_i } \right\rangle \to \left| {c_i } \right\rangle ;1 \le i \le n - 1\\ 
 \left| t \right\rangle \to \left| t_{CU}\right\rangle = U^{c_1 c_2 ...c_{n - 1} }\left| t \right\rangle 
\\ 
 \end{array}
\end{equation}

\noindent
where $c_1 c_2 ...c_{n - 1}$ in the exponent of $U$ means the $AND$-ing of the 
qubits $c_1 ,\,c_2 ,...,c_{n - 1} $.


\begin{center}
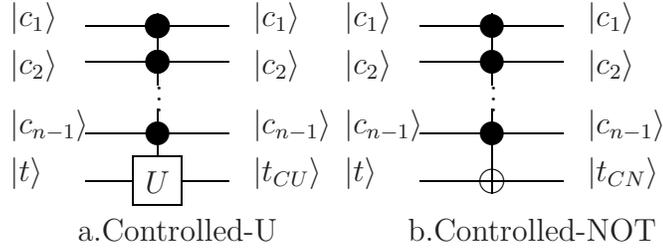
\begin{figure} [H]
\begin{center}
\setlength{\unitlength}{3947sp}%
\begingroup\makeatletter\ifx\SetFigFont\undefined%
\gdef\SetFigFont#1#2#3#4#5{%
  \reset@font\fontsize{#1}{#2pt}%
  \fontfamily{#3}\fontseries{#4}\fontshape{#5}%
  \selectfont}%
\fi\endgroup%
\begin{picture}(3525,1485)(3376,-1936)
{\color[rgb]{0,0,0}\thinlines
\put(4201,-1261){\circle*{150}}
}%
{\color[rgb]{0,0,0}\put(4201,-586){\circle*{150}}
}%
{\color[rgb]{0,0,0}\put(4201,-811){\circle*{150}}
}%
{\color[rgb]{0,0,0}\put(6301,-586){\circle*{150}}
}%
{\color[rgb]{0,0,0}\put(6301,-811){\circle*{150}}
}%
{\color[rgb]{0,0,0}\put(6301,-1261){\circle*{150}}
}%
{\color[rgb]{0,0,0}\put(6301,-1561){\circle{150}}
}%
{\color[rgb]{0,0,0}\put(4201,-811){\circle*{150}}
}%
{\color[rgb]{0,0,0}\put(4201,-586){\circle*{150}}
}%
{\color[rgb]{0,0,0}\put(4201,-586){\circle{150}}
}%
{\color[rgb]{0,0,0}\put(4051,-1711){\framebox(300,300){}}
}%
{\color[rgb]{0,0,0}\put(4351,-1561){\line( 1, 0){300}}
}%
{\color[rgb]{0,0,0}\put(4051,-1561){\line(-1, 0){300}}
}%
{\color[rgb]{0,0,0}\put(4201,-1111){\line( 0,-1){300}}
}%
{\color[rgb]{0,0,0}\put(4651,-586){\line(-1, 0){900}}
}%
{\color[rgb]{0,0,0}\put(4201,-511){\line( 0,-1){450}}
}%
{\color[rgb]{0,0,0}\put(4651,-811){\line(-1, 0){900}}
}%
{\color[rgb]{0,0,0}\put(4651,-1261){\line(-1, 0){900}}
}%
{\color[rgb]{0,0,0}\put(6751,-586){\line(-1, 0){900}}
}%
{\color[rgb]{0,0,0}\put(6751,-811){\line(-1, 0){900}}
}%
{\color[rgb]{0,0,0}\put(6751,-1261){\line(-1, 0){900}}
}%
{\color[rgb]{0,0,0}\put(6751,-1561){\line(-1, 0){900}}
}%
{\color[rgb]{0,0,0}\put(6301,-511){\line( 0,-1){450}}
}%
{\color[rgb]{0,0,0}\put(6301,-1111){\line( 0,-1){525}}
}%

\put(4185,-1120){$\vdots$}
\put(6285,-1120){$\vdots$}

\put(4801,-586){$\left| {c_1 } \right\rangle$}%
\put(3276,-586){$\left| {c_1 } \right\rangle$}%
\put(5376,-586){$\left| {c_1 } \right\rangle$}%
\put(6901,-586){$\left| {c_1 } \right\rangle$}%

\put(4801,-886){$\left| {c_2 } \right\rangle$}%
\put(3276,-886){$\left| {c_2 } \right\rangle$}%
\put(5376,-886){$\left| {c_2 } \right\rangle$}%
\put(6901,-886){$\left| {c_2 } \right\rangle$}%

\put(4801,-1561){$\left| {t_{CU} } \right\rangle$}%
\put(3276,-1561){$\left| {t } \right\rangle$}%
\put(5376,-1561){$\left| {t } \right\rangle$}%
\put(6901,-1561){$\left| {t_{CN} } \right\rangle$}%

\put(4801,-1261){$\left| {c_{n-1} } \right\rangle$}%
\put(3276,-1261){$\left| {c_{n-1} } \right\rangle$}%
\put(5376,-1261){$\left| {c_{n-1} } \right\rangle$}%
\put(6901,-1261){$\left| {c_{n-1} } \right\rangle$}%

\put(4126,-1636){$U$}%

\put(5776,-1936){b.Controlled-NOT}%
\put(3700,-1936){a.Controlled-U}%
\end{picture}
\end{center}
\caption{Controlled gates where the back circle $\bullet $ indicates 
the control qubits, and the symbol $ \oplus $ in part (b.) indicates the target qubit.}
\label{ENhfig3_4_X}
\end{figure}
\end{center}

If $U$ in the general Controlled-$U$ gate is replaced with the $X$ gate mentioned above, the resulting gate is  
called Controlled-$NOT$ gate (shown in Fig.(\ref{ENhfig3_4_X}.b)). It works as follows: It 
inverts the target qubit if and only if all the control qubits are set to 1. 
Thus the qubits of the system will be transformed 
according to the following rule:

\begin{equation}
\label{ENheq8}
\begin{array}{l}
 \left| {c_i } \right\rangle \to \left| {c_i } \right\rangle ;1 \le i \le n - 1 \\ 
 \left| {t } \right\rangle \to \left| t_{CN}\right\rangle = \left| {t \oplus c_1 c_2 ...c_{n - 1} } 
\right\rangle \\ 
 \end{array}
\end{equation}

\noindent
where $c_1 c_2 \ldots c_{n - 1} $ is the $AND$-ing of the qubits $c_1 ,c_2 
,\ldots ,c_{n - 1} $ and $ \oplus $ is the classical XOR operation.

\section{Search Problem}

Consider a list $L$ of $N$ items; $L = \{ 0,1,...,N - 1\}$, and consider a function $f$ 
which maps the items in $L$ to either 0 or 1 according to some properties these items shall satisfy; 
i.e. $f:L \to \{ 0,1\}$. The problem is to find any $i \in L$ such that $f(i) = 1$ assuming that 
such $i$ must exist in the list. It was shown classically that we need approximately 
${N \mathord{\left/ {\vphantom {N 2}} \right. \kern-\nulldelimiterspace} 2}$ tests to get a result with 
probability at least one-half. Let $M$ denotes the number of matches within the search space such that 
$1 \le M \le N$ and for simplicity and without loss of generality we can assume that $N = 2^n$. 
Grover's algorithm was shown to solve this problem \cite{boyer96} in $O\left( {\sqrt {N/M} } \right)$. 
In \cite{niel00}, it was shown that the number of iterations will increase 
for $M > N/2$ which is undesired behaviour for a search algorithm. To over come this problem it was proposed 
in \cite{niel00} that the search space can be doubled so the number of matches always less than half the search space and iterate 
the algorithm $\pi /4\sqrt {2N/M}$ times so the algorithm still runs in $O\left( {\sqrt {N/M} } \right)$. 
But using this approach will double the cost of space/time requirement. In the following section we will present 
an algorithm that can find a solution for $M > N/2$ with probability at least $92.6\%$ after applying the algorithm 
once.

\section{The Algorithm}
\subsection{Iterating the algorithm once}

\begin{center}
\begin{figure} 
\begin{center}
\setlength{\unitlength}{3947sp}%
\begingroup\makeatletter\ifx\SetFigFont\undefined%
\gdef\SetFigFont#1#2#3#4#5{%
  \reset@font\fontsize{#1}{#2pt}%
  \fontfamily{#3}\fontseries{#4}\fontshape{#5}%
  \selectfont}%
\fi\endgroup%
\begin{picture}(4050,2091)(76,-1690)
\thinlines
{\color[rgb]{0,0,0}\put(1501, 89){\framebox(300,300){}}
}%
{\color[rgb]{0,0,0}\put(1501,-286){\framebox(300,300){}}
}%
{\color[rgb]{0,0,0}\put(1501,-886){\framebox(300,300){}}
}%
{\color[rgb]{0,0,0}\put(1951,-1261){\framebox(750,1650){}}
}%
{\color[rgb]{0,0,0}\put(2851,-1261){\framebox(750,1650){}}
}%
{\color[rgb]{0,0,0}\put(3976,-886){\line( 0, 1){1200}}
}%
{\color[rgb]{0,0,0}\put(3976,-886){\line(-1, 0){ 75}}
}%
{\color[rgb]{0,0,0}\put(3901,314){\line( 1, 0){ 75}}
}%
{\color[rgb]{0,0,0}\put(1876,-1336){\line( 0,-1){ 75}}
}%
{\color[rgb]{0,0,0}\put(1876,-1411){\line( 1, 0){1875}}
}%
{\color[rgb]{0,0,0}\put(3751,-1411){\line( 0, 1){ 75}}
}%
{\color[rgb]{0,0,0}\put(826,-886){\line( 0, 1){1200}}
}%
{\color[rgb]{0,0,0}\put(901,-886){\line(-1, 0){ 75}}
}%
{\color[rgb]{0,0,0}\put(826,314){\line( 1, 0){ 75}}
}%
{\color[rgb]{0,0,0}\put(1201,239){\line( 1, 0){300}}
}%
{\color[rgb]{0,0,0}\put(1801,239){\line( 1, 0){150}}
}%
{\color[rgb]{0,0,0}\put(2701,239){\line( 1, 0){150}}
}%
{\color[rgb]{0,0,0}\put(3601,239){\line( 1, 0){225}}
}%
{\color[rgb]{0,0,0}\put(1201,-136){\line( 1, 0){300}}
}%
{\color[rgb]{0,0,0}\put(1801,-136){\line( 1, 0){150}}
}%
{\color[rgb]{0,0,0}\put(2701,-136){\line( 1, 0){150}}
}%
{\color[rgb]{0,0,0}\put(3601,-136){\line( 1, 0){225}}
}%
{\color[rgb]{0,0,0}\put(1201,-736){\line( 1, 0){300}}
}%
{\color[rgb]{0,0,0}\put(1801,-736){\line( 1, 0){150}}
}%
{\color[rgb]{0,0,0}\put(2701,-736){\line( 1, 0){150}}
}%
{\color[rgb]{0,0,0}\put(3601,-736){\line( 1, 0){225}}
}%
{\color[rgb]{0,0,0}\put(1201,-1111){\line( 1, 0){750}}
}%
{\color[rgb]{0,0,0}\put(2701,-1111){\line( 1, 0){150}}
}%
{\color[rgb]{0,0,0}\put(3601,-1111){\line( 1, 0){225}}
}%

\put(1651,-500){$\vdots$}

\put(3751,-500){$\vdots$}

\put(301, 14){$n$}%

\put(151,-211){qubits}%

\put( 76,-1261){workspace}%

\put(151,-1036){1 qubit}%

\put(4126,-211){Measure}%

\put(980,239){$\left|0\right\rangle$}%
\put(980,-136){$\left|0\right\rangle$}%
\put(980,-736){$\left|0\right\rangle$}%
\put(980,-1111){$\left|0\right\rangle$}%

\put(1576,164){$H$}%
\put(1576,-211){$H$}%
\put(1576,-811){$H$}%

\put(3151,-436){$Y$}%

\put(2251,-436){$U_f$}%

\put(2251,-1636){$O\left( {\sqrt {N/M} } \right)$}

\end{picture}

\end{center}
\caption{Quantum circuit for the proposed algorithm.}
\label{ENhfig6}
\end{figure}
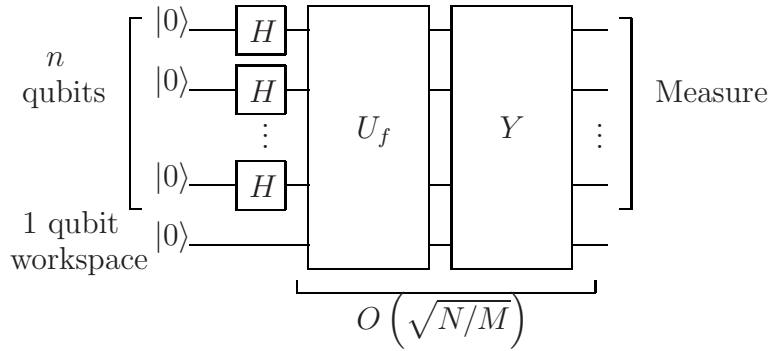
\end{center}

For a list of size $N=2^n$, the steps of the algorithm can be understood 
as follows as shown in Fig.(\ref{ENhfig6}): 

\begin{itemize}
\item[1-]{\it Register Preparation}. Prepare a quantum register of $n+1$ 
qubits all in state $\left| 0 \right\rangle $, where the extra qubit is 
used as a workspace for evaluating the oracle $U_f$:

\begin{equation}
\label{ENheq10}
\left| {W_0 } \right\rangle = \left| 0 \right\rangle ^{ \otimes n} \otimes 
\left| 0 \right\rangle. 
\end{equation}

\item[2-] {\it Register Initialization}. Apply Hadamard gate on each of the first $n$ qubits in parallel, 
so they contain the $2^{n}$ states, where $i$ is the integer representation of items in the list:

\begin{equation}
\label{ENheq11}
\left| {W_1 } \right\rangle = \left( {H^{ \otimes n} \otimes I} 
\right)\left| {W_0 } \right\rangle = \left( {\frac{1}{\sqrt N 
}\sum\limits_{i = 0}^{N - 1} {\left| i \right\rangle } } \right) \otimes 
\left| 0 \right\rangle.
\end{equation}

\item[3-] {\it Applying Oracle}. Apply the oracle $U_{f}$ to map the items in the list to either 0 or 1
simultaneously and stores the result in the extra workspace qubit:

\begin{equation}
\label{ENheq12}
\left| {W_2 } \right\rangle = 
\frac{1}{\sqrt N }\sum\limits_{i = 0}^{N - 1} {\left( {\left| i 
\right\rangle \otimes \left| {0 \oplus f(i)} \right\rangle } \right)} = 
\frac{1}{\sqrt N }\sum\limits_{i = 0}^{N - 1} {\left( {\left| i 
\right\rangle \otimes \left| {f(i)} \right\rangle } \right)}. 
\end{equation}


\item[4-]{\it Partial Diffusion}. 
In this step, we will define a new operator: {\it Partial Diffusion Operator} $(Y)$ which works similar to the 
well known {\it Diffusion Operator} used in Grover's algorithm \cite{grover96} except that it performs the 
{\it inversion about the mean} operation only on a subspace of the system as follows: The diagonal representation 
of the partial diffusion operator $Y$ when applied on $n+1$ qubits system can take this form:

\begin{equation}
\label{ENheq13}
Y = H^{ \otimes n}  \otimes I\left( {2\left| 0 \right\rangle \left\langle 0 \right| - I} \right)H^{ \otimes n}  \otimes I,
\end{equation}

where the vector $\left| 0 \right\rangle$ used in Eqn.(\ref{ENheq13}) is a vector of lenght $P = 2N = 2^{n+1}$. 
Applying $Y$ on a general system $\sum\nolimits_{k = 0}^{P - 1} {\delta _k \left| k \right\rangle }$; 
where, $\left| {\delta _k } \right|^2  = 1$, can be understood as follows: Without loosing of generality, the general system can be 
re-written as, 

\begin{equation}
\label{ENheq14}
\sum\limits_{k = 0}^{P - 1} {\delta _k \left| k \right\rangle }=\sum\limits_{j = 0}^{N - 1} {\alpha _j \left( {\left| j \right\rangle  \otimes \left| 0 \right\rangle } \right)}  + \sum\limits_{j = 0}^{N - 1} {\beta _j \left( {\left| j \right\rangle  \otimes \left| 1 \right\rangle } \right)}, 
\end{equation}

\noindent
where \{$\alpha _j  = \delta _k$ : $k$ even\} and \{$\beta _j  = \delta _k$ : $k$ odd\}, 
then applying $Y$ on the system gives,

\begin{equation}
\label{ENheq15}
\begin{array}{l}
Y\left( {\sum\limits_{k = 0}^{P - 1} {\delta _k \left| k \right\rangle } } \right) = \left( {H^{ \otimes n}  \otimes I\left( {2\left| 0 \right\rangle \left\langle 0 \right| - I} \right)H^{ \otimes n}  \otimes I} \right)\sum\limits_{k = 0}^{P - 1} {\delta _k \left| k \right\rangle } \\
\,\,\,\,\,\,\,\,\,\,\,\,\,\,\,\,\,\,\,\,\,\,\,\,\,\,\,\,\,\,\,\,\,\,\,\,\,\,\, = 2\left( {H^{ \otimes n}  \otimes I\left| 0 \right\rangle \left\langle 0 \right|H^{ \otimes n}  \otimes I} \right)\sum\limits_{k = 0}^{P - 1} {\delta _k \left| k \right\rangle }  - \sum\limits_{k = 0}^{P - 1} {\delta _k \left| k \right\rangle } \\
\,\,\,\,\,\,\,\,\,\,\,\,\,\,\,\,\,\,\,\,\,\,\,\,\,\,\,\,\,\,\,\,\,\,\,\,\,\,\, =\sum\limits_{j = 0}^{N - 1} {\left( {2\left\langle \alpha  \right\rangle  - \alpha _j } \right)\left( {\left| j \right\rangle  \otimes \left| 0 \right\rangle } \right)}  - \sum\limits_{j = 0}^{N - 1} {\beta _j \left( {\left| j \right\rangle  \otimes \left| 1 \right\rangle } \right)}, \\
\end{array}
\end{equation}

\noindent
where $\left\langle \alpha  \right\rangle  = \frac{1}{N}\sum\nolimits_{j = 0}^{N - 1} {\alpha _j }$ is the mean of 
the amplitudes of the subspace $\sum\nolimits_{j = 0}^{N - 1} {\alpha _j \left( {\left| j \right\rangle  \otimes 
\left| 0 \right\rangle } \right)}$; i.e. applying the operator $Y$ will perform the inversion about the mean 
only on the subspace; $\sum\nolimits_{j = 0}^{N - 1} {\alpha _j \left( {\left| j \right\rangle  \otimes 
\left| 0 \right\rangle } \right)}$ and will only change the sign of the amplitudes for the rest of the system; 
$\sum\nolimits_{j = 0}^{N - 1} {\beta _j \left( {\left| j \right\rangle  \otimes \left| 1 \right\rangle } \right)}$, 
a circuit implementation using elementary gates \cite{bare95} is shown in Fig.(\ref{figY}).

\begin{center}
\begin{figure} 
\begin{center}
\setlength{\unitlength}{3947sp}%
\begingroup\makeatletter\ifx\SetFigFont\undefined%
\gdef\SetFigFont#1#2#3#4#5{%
  \reset@font\fontsize{#1}{#2pt}%
  \fontfamily{#3}\fontseries{#4}\fontshape{#5}%
  \selectfont}%
\fi\endgroup%
\begin{picture}(3525,1974)(3226,-2173)
{\color[rgb]{0,0,0}\thinlines
\put(5101,-361){\circle*{150}}
}%
{\color[rgb]{0,0,0}\put(5101,-811){\circle*{150}}
}%
{\color[rgb]{0,0,0}\put(5101,-1561){\circle*{150}}
}%
{\color[rgb]{0,0,0}\put(3976,-511){\framebox(300,300){}}
}%
{\color[rgb]{0,0,0}\put(4276,-361){\line( 1, 0){225}}
}%
{\color[rgb]{0,0,0}\put(4501,-511){\framebox(300,300){}}
}%
{\color[rgb]{0,0,0}\put(3976,-961){\framebox(300,300){}}
}%
{\color[rgb]{0,0,0}\put(4276,-811){\line( 1, 0){225}}
}%
{\color[rgb]{0,0,0}\put(4501,-961){\framebox(300,300){}}
}%
{\color[rgb]{0,0,0}\put(3976,-1711){\framebox(300,300){}}
}%
{\color[rgb]{0,0,0}\put(4276,-1561){\line( 1, 0){225}}
}%
{\color[rgb]{0,0,0}\put(4501,-1711){\framebox(300,300){}}
}%
{\color[rgb]{0,0,0}\put(4801,-361){\line( 1, 0){600}}
}%
{\color[rgb]{0,0,0}\put(4801,-811){\line( 1, 0){600}}
}%
{\color[rgb]{0,0,0}\put(4801,-1561){\line( 1, 0){600}}
}%
{\color[rgb]{0,0,0}\put(4951,-2161){\framebox(300,300){}}
}%
{\color[rgb]{0,0,0}\put(5101,-361){\line( 0,-1){600}}
}%
{\color[rgb]{0,0,0}\put(5101,-1411){\line( 0,-1){450}}
}%
{\color[rgb]{0,0,0}\put(5401,-511){\framebox(300,300){}}
}%
{\color[rgb]{0,0,0}\put(5401,-961){\framebox(300,300){}}
}%
{\color[rgb]{0,0,0}\put(5401,-1711){\framebox(300,300){}}
}%
{\color[rgb]{0,0,0}\put(5401,-2161){\framebox(300,300){}}
}%
{\color[rgb]{0,0,0}\put(5926,-511){\framebox(300,300){}}
}%
{\color[rgb]{0,0,0}\put(5926,-961){\framebox(300,300){}}
}%
{\color[rgb]{0,0,0}\put(5926,-1711){\framebox(300,300){}}
}%
{\color[rgb]{0,0,0}\put(5701,-1561){\line( 1, 0){225}}
}%
{\color[rgb]{0,0,0}\put(6226,-1561){\line( 1, 0){ 75}}
}%
{\color[rgb]{0,0,0}\put(5701,-811){\line( 1, 0){225}}
}%
{\color[rgb]{0,0,0}\put(5701,-361){\line( 1, 0){225}}
}%
{\color[rgb]{0,0,0}\put(6226,-811){\line( 1, 0){ 75}}
}%
{\color[rgb]{0,0,0}\put(6226,-361){\line( 1, 0){ 75}}
}%
{\color[rgb]{0,0,0}\put(3901,-361){\line( 1, 0){ 75}}
}%
{\color[rgb]{0,0,0}\put(3901,-811){\line( 1, 0){ 75}}
}%
{\color[rgb]{0,0,0}\put(3901,-1561){\line( 1, 0){ 75}}
}%
{\color[rgb]{0,0,0}\put(4951,-2011){\line(-1, 0){1050}}
}%
{\color[rgb]{0,0,0}\put(5251,-2011){\line( 1, 0){150}}
}%
{\color[rgb]{0,0,0}\put(5701,-2011){\line( 1, 0){600}}
}%
{\color[rgb]{0,0,0}\put(3826,-286){\line( 0,-1){1350}}
}%
{\color[rgb]{0,0,0}\put(3826,-1636){\line( 1, 0){ 75}}
}%
{\color[rgb]{0,0,0}\put(3826,-286){\line( 1, 0){ 75}}
}%

\put(4105,-1236){$\vdots$}
\put(4630,-1236){$\vdots$}
\put(5080,-1236){$\vdots$}%
\put(5530,-1236){$\vdots$}%
\put(6055,-1236){$\vdots$}%

\put(4051,-436){$H$}%
\put(4051,-886){$H$}%
\put(4051,-1636){$H$}%
\put(4576,-436){$X$}%
\put(4576,-886){$X$}%
\put(4576,-1636){$X$}%
\put(5476,-436){$X$}%
\put(5476,-886){$X$}%
\put(5476,-1636){$X$}%
\put(5026,-2086){$U$}%
\put(5476,-2086){$V$}%
\put(6001,-436){$H$}%
\put(6001,-886){$H$}%
\put(6001,-1636){$H$}%
\put(6576,-511){$U = \left[ {\begin{array}{*{20}c}
   { - 1} & 0  \\
   0 & 1  \\
\end{array}} \right]$}%
\put(6576,-1636){$V = \left[ {\begin{array}{*{20}c}
   { - 1} & 0  \\
   0 & { - 1}  \\
\end{array}} \right]$}
\put(3376,-736){$n$}%
\put(3226,-961){qubits}%
\end{picture}
\end{center}
\caption{Quantum circuit representing the Partial Diffusion Operator $Y$ over $n+1$ qubits.}
\label{figY}
\end{figure}
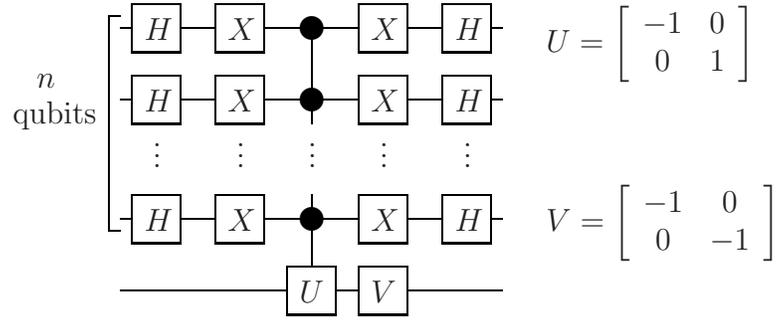
\end{center}

The main idea of using the partial diffusion operator in searching is to apply the inversion about the mean operation 
only on the subspace of the system which includes all the states which represent the non-matches and half the number of 
the states which represent the matches while the other half will have the sign of their amplitudes inverted to the 
negative sign preparing them to be involved in the partial diffusion operation in the next iteration so the amplitudes 
of the matching states get amplified partially each iteration. The benefit of this is to keep half the number of 
the states which represent the matches as a stock each iteration to resist the {\it de-amplification behaviour} of the 
diffusion operation when reaching the turning points as we will see when examining the performance of the algorithm.

Let $M$ be the number of matches, which makes the oracle $U_f$ evaluate to TRUE (solutions); such that 
$1 \le M \le N$; assume that $\sum\nolimits_i {{'}} $ indicates a sum over all $i$ which are desired matches 
($M$ states), and $\sum\nolimits_i {{''}} $ indicates a sum over all $i$ which are undesired items in the list. 
So, the system  $\left| {W_2} \right\rangle$ shown in Eqn.(\ref{ENheq12}) can be written as follows:

\begin{equation}
\label{ENheq16}
\left| {W_2} \right\rangle  = \frac{1}{{\sqrt N }}\sum\limits_{i = 0}^{N - 1} {''\left( {\left| i \right\rangle  \otimes \left| 0 \right\rangle } \right)}  + \frac{1}{{\sqrt N }}\sum\limits_{i = 0}^{N - 1} {'\left( {\left| i \right\rangle  \otimes \left| 1 \right\rangle } \right)}. 
\end{equation}

Applying $Y$ on $\left| {W_2} \right\rangle$ will result in a new system described as follows:

\begin{equation}
\label{ENheq17}
\left| {W_3} \right\rangle  = a_1 \sum\limits_{i = 0}^{N - 1} {''\left( {\left| i \right\rangle  \otimes \left| 0 \right\rangle } \right)}  + b_1 \sum\limits_{i = 0}^{N - 1} {'\left( {\left| i \right\rangle  \otimes \left| 0 \right\rangle } \right)}  + c_1 \sum\limits_{i = 0}^{N - 1} {'\left( {\left| i \right\rangle  \otimes \left| 1 \right\rangle } \right)}, 
\end{equation}

\noindent
where the mean used in the definition of partial diffusion operator is,

\begin{equation}
\label{ENheq18}
\left\langle {\alpha _1 } \right\rangle  = \left( {\frac{{N - M}}{{N\sqrt N }}} \right),
\end{equation}

and $a_1$, $b_1$ and $c_1$ used in Eqn.(\ref{ENheq17}) are calculates as follows:

\begin{equation}
\label{ENheq19}
a_1  = 2\left\langle {\alpha _1 } \right\rangle  - \frac{1}{{\sqrt N }};\,\,\,\,\,\
b_1  = 2\left\langle {\alpha _1 } \right\rangle; \,\,\,\,\,\ 
c_1  = \frac{{ - 1}}{{\sqrt N }}.
\end{equation}

Such that,

\begin{equation}
\label{ENheq20}
\left( {N - M} \right)a_1^2  + Mb_1^2  + Mc_1^2  = 1.
\end{equation}

Notice that, the states with amplitude $b_1$ was with amplitude {\it zero} before applying $Y$.

\item[5-] {\it Measurement}. If we measure the first $n$ qubits after the first iteration ($q=1$), 
we will get the desired solution with probability given as follows:

\begin{itemize}
\item[i-]Probability $P_{s}^{(1)}$ to find a match out of the $M$ possible matches; taking into account that a 
solution $\left| i \right\rangle $ occurs {\it twice} as: $\left( {\left| i \right\rangle \otimes \left| 0 
\right\rangle } \right)$ with amplitude $b_1$ and $\left( {\left| i \right\rangle \otimes \left| 1 \right\rangle 
} \right)$ with amplitude $c_1$ as shown in Eqn.(\ref{ENheq17}), can be calculated as follows:

\begin{equation}
\label{ENheq21}
\begin{array}{l}
 P_{s}^{(1)}  = M\left( {b_1^2  + c_1^2 } \right) \\ 
\,\,\,\,\,\,\,  = M\left( {\left( {\frac{{2\left( {N - M} \right)}}{{N\sqrt N }}} \right)^2  + \left( {\frac{{ - 1}}{{\sqrt N }}} \right)^2 } \right) \\ 
\,\,\,\,\,\,\,  = 5\left( {\frac{M}{N}} \right) - 8\left( {\frac{M}{N}} \right)^2  + 4\left( {\frac{M}{N}} \right)^3.  \\ 
 \end{array}
\end{equation}

\item[ii-] Probability $P_{ns}^{(1)}$ to find undesired result out of the states can be calculated as follows:

\begin{equation}
\label{ENheq22}
P_{ns}^{(1)}  = (N - M)a_1^2. 
\end{equation}

Notice that, using Eqn.(\ref{ENheq20}), 

\begin{equation}
\label{ENheq23}
P_{s}^{(1)}  + P_{ns}^{(1)}  = 1.
\end{equation}
\end{itemize}
\end{itemize}
\subsubsection{Performance after Iterating the Algorithm Once}

\begin{table}[H]
\begin{center}
\begin{tabular}
{|c|c|c|c|}
\hline
 $n$, where $N=2^n$ & 
\ Max. prob.  & 
 Min. prob.  & 
 Avg. prob.   \\ \hline
2  & 
1.0  & 
0.8125  & 
0.875  \\ \hline
 3  & 
 1.0  & 
 0.507812  & 
 0.937500  \\ \hline
 4  & 
 1.0  & 
 0.282227  & 
0.968750 \\ \hline
 5  & 
 1.0  & 
 0.148560  & 
 0.984375 \\ \hline
 6  & 
 1.0  & 
 0.076187  & 
 0.992187  \\ \hline

\end{tabular}
\caption {First iteration performance with different size search space.}
\label{ENhtab3}
\end{center}
\end{table}

Considering Eqn.(\ref{ENheq17}) and Eqn.(\ref{ENheq21}) we can see that 
the probability to find a solution varies according to the number of matches $M$ in the superposition. 

From Table.\ref{ENhtab3}, we can see that the maximum probability is always 1.0,
and the minimum probability (worst case) decreases as the size of the list increases, 
which is expected for small $M$ because the number of states will increase and 
the probability shall distribute over more states while the average 
probability increases as the size of the list increases. It implies 
that the average performance of the first iteration of algorithm to find a solution increases as the 
size of the list increases.

To verify these results, taking into account that the oracle $U_f$ is taken as a black box, we can define 
the average probability of success of the first iteration of algorithm; $average(P_{s}^{(1)})$, as follows:

\begin{equation}
\label{ENheq24}
\begin{array}{l}
 average(P_{s}^{(1)} ) = \frac{1}{{2^N }}\sum\limits_{M = 1}^N {{}^NC_M P_{s}^{(1)} }  \\ 
 \,\,\,\,\,\,\,\,\,\,\,\,\,\,\,\,\,\,\,\,\,\,\,\,\,\, = \frac{1}{{2^N }}\sum\limits_{M = 1}^N {\frac{{N!}}{{M!(N - M)!}}.M\left( {b_1^2  + c_1^2 } \right)}  \\ 
 \,\,\,\,\,\,\,\,\,\,\,\,\,\,\,\,\,\,\,\,\,\,\,\,\,\, = \frac{1}{{2^{N + 1} N^3 }}\sum\limits_{M = 1}^N {\frac{{N!}}{{(M - 1)!(N - M)!}}\left( {10N^2  - 16MN + 8M^2 } \right)}  \\ 
 \,\,\,\,\,\,\,\,\,\,\,\,\,\,\,\,\,\,\,\,\,\,\,\,\,\, = 1-\frac{1}{{2N}}.\\
 \end{array}
\end{equation}

\noindent
where ${}^NC_M  = \frac{{N!}}{{M!(N - M)!}}$ is the number of possible cases for $M$ matches.



\noindent
We can see that as the size of the list increases $(N \to \infty)$, $average (P_{s}^{(1)})$ shown in 
Eqn.(\ref{ENheq24}) tends to $1$.


Classically, we can try to find a random guess of the item, 
which represents the solution (one trial guess), we may succeed to find a solution with probability 
$P^{(classical)}_{s} = M/N$. The average probability of success can be calculated as follows:

\begin{equation}
\label{ENheqn26}
\begin{array}{l}
average(P_s^{(classical)} ) = \frac{1}{{2^N }}\sum\limits_{M = 1}^N {{}^NC_M P_s^{(classical)} } \\
\,\,\,\,\,\,\,\,\,\,\,\,\,\,\,\,\,\,\,\,\,\,\,\,\,\,\,\,\,\,\,\,\,\,\,\,\,\,\,\,\,\,\,\,\,\,\,\,\,\,\,\,\,\,\,= \frac{1}{{2^N }}\sum\limits_{M = 1}^N {\frac{{N.M}}{{M!(N - M)!N}}} \\
\,\,\,\,\,\,\,\,\,\,\,\,\,\,\,\,\,\,\,\,\,\,\,\,\,\,\,\,\,\,\,\,\,\,\,\,\,\,\,\,\,\,\,\,\,\,\,\,\,\,\,\,\,\,\,= \frac{1}{2}. \\ 
\end{array}
\end{equation}

It means that we have an average probability one-half to find or not to find a solution by a single random guess 
even with the increase in size of the list.

Similarly, Grover's algorithm has an average probabilty {\it one-half} after arbitrary number of iterations 
as we will see. It was shown in \cite{boyer96} that the probability of success of Grover's algorithm after $q_G$ 
iterations is given by:

\begin{equation}
\label{ENheqn27}
 P_s^{(q_G )} = \sin ^2 ((2q_G + 1)\theta ), \,\,\, \mbox{where, }\,0 < \theta  \le \frac{\pi }{2}\mbox{ and }\sin ^2 (\theta ) = \frac{M}{N}.\\ 
\end{equation}

The average probability of success of Grover's algorithm after arbitrary number of iterations 
is as follows (Appendix A in \cite{younes03}):

\begin{equation}
\label{ENheqn28}
 average(P_s^{(q_G )}) = \frac{1}{{2^N }}\sum\limits_{M = 1}^N {{}^NC_M \sin ^2 ((2q_G + 1)\theta )}  = \frac{1}{2}. \\ 
\end{equation}

\begin{figure}[htbp]
\centerline{\includegraphics[width=4.02in,height=3.403in]{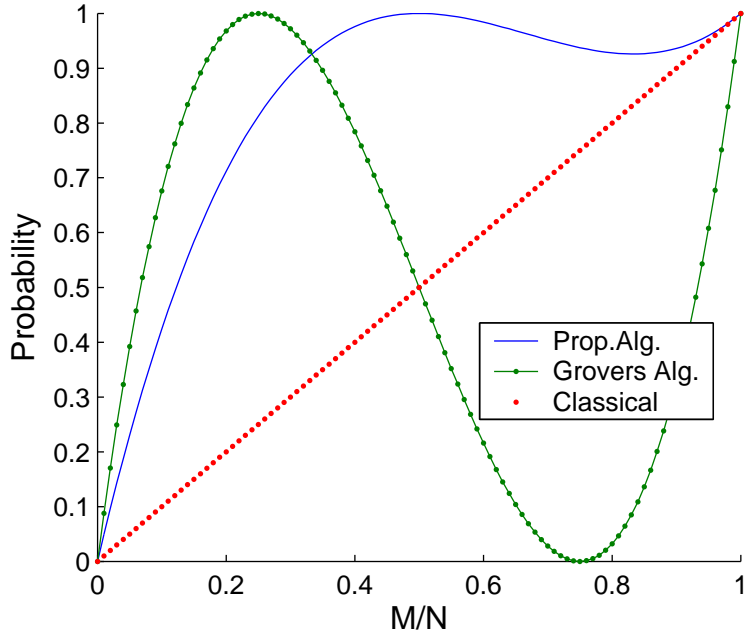}}
\caption{A plot of the probability of success of the first the iteration of proposed algorithm $P_s^{(1)}$, first iteration of Grover's algorithm 
$P_s^{(1_G )}$ and the classical guess $P_s^{(classical)}$ as a function of the ratio $(M/N)$.} 
\label{ENhplot}
\end{figure}

Comparing the performance of the first iteration of the proposed algorithm, 
first iteration of Grover's algorithm and the classical guess 
technique, Fig.(\ref{ENhplot}) shows the probability of success of the three algorithms just mentioned as a function of 
the ratio $(M/N)$.


We can see from Fig.(\ref{ENhplot}) that the probability of success of the first iteration is always above 
that of the classical guess technique. Grover's algorithm solves the case where $M=N/4$ with certainity and the proposed 
algorithm solves the case where $M=N/2$ with certainity. The probability of success of Grover's algorithm 
will start to go below one-half for $M>N/2$ while the probability of success of the proposed algorithm will stay 
more reliable with propabilty at least $92.6\%$.

\subsection{Iterating the Algorithm}

If we consider iterating the algorithm, the iterating block will be applying the oracle $U_{f}$ and the operator $Y$ on 
the system in sequence as 
shown in Fig.(\ref{ENhfig6}). To understand the effect of each iteration on the system, 
we will trace the state of the system during the first few iterations. 
Consider the system after first iteration shown in Eqn.(\ref{ENheq17}), second iteration 
will modify the system as follows:

Apply the oracle $U_{f}$ will {\it swap} the amplitudes of 
the states which represent only the matches, i.e. states with amplitudes 
$b_{1}$ will be with amplitudes $c_{1}$ and states with amplitudes $c_{1}$ will 
be with amplitudes $b_{1}$ so the system can be described as,

\begin{equation}
\label{ENheqn29}
\left| {W_4 } \right\rangle = a_1 \sum\limits_{i = 0}^{N - 1} {''\left( 
{\left| i \right\rangle \otimes \left| 0 \right\rangle } \right)} + c_1 
\sum\limits_{i = 0}^{N - 1} {'\left( {\left| i \right\rangle \otimes \left| 
0 \right\rangle } \right)} + b_1 \sum\limits_{i = 0}^{N - 1} {'\left( 
{\left| i \right\rangle \otimes \left| 1 \right\rangle } \right)} .
\end{equation}

Applying the operator $Y$ will change the system as follows,

\begin{equation}
\label{ENheqn30}
\left| {W_5 } \right\rangle = a_2 \sum\limits_{i = 0}^{N - 1} {''\left( 
{\left| i \right\rangle \otimes \left| 0 \right\rangle } \right)} + b_2 
\sum\limits_{i = 0}^{N - 1} {'\left( {\left| i \right\rangle \otimes \left| 
0 \right\rangle } \right)} + c_2 \sum\limits_{i = 0}^{N - 1} {'\left( 
{\left| i \right\rangle \otimes \left| 1 \right\rangle } \right)} ,
\end{equation}

\noindent
where the mean used in the definition of partial diffusion operator is,

\begin{equation}
\label{ENheqn31}
\left\langle {\alpha _2 } \right\rangle = \frac{1}{N}\left( {\left( {N - M} 
\right)a_1 + Mc_1 } \right),
\end{equation}

\noindent
and $a_{2}$, $b_{2}$ and $c_{2}$ used in Eqn.(\ref{ENheqn30}) are calculated as follows:

\begin{equation}
\label{ENheqn32}
a_2 = 2\left\langle {\alpha _2 } \right\rangle - a_1 ;\,\,\,\,\,\,\,\,b_2 = 
2\left\langle {\alpha _2 } \right\rangle - c_1 ;\,\,\,\,\,\,c_2 = - b_1 ,
\end{equation}

\noindent
and the probabilities of the system are,

\begin{equation}
\label{ENheqn33}
\begin{array}{l}
 P_s^{(2)} = M\left( {b_2^2 + c_2^2 } \right) = M\left( {b_2^2 + b_1^2 } 
\right). \\ 
 P_{ns}^{(2)} = M\left( {a_2^2 } \right) = M\left( {b_2 + c_2 } \right)^2 = 
M\left( {b_2 - b_1 } \right)^2. \\ 
 \end{array}
\end{equation}

In the same fashion, third iteration will give the following system,

\begin{equation}
\label{ENheqn34}
U_f \left| {W_5 } \right\rangle = \left| {W_6 } \right\rangle = a_2 
\sum\limits_{i = 0}^{N - 1} {''\left( {\left| i \right\rangle \otimes \left| 
0 \right\rangle } \right)} + c_2 \sum\limits_{i = 0}^{N - 1} {'\left( 
{\left| i \right\rangle \otimes \left| 0 \right\rangle } \right)} + b_2 
\sum\limits_{i = 0}^{N - 1} {'\left( {\left| i \right\rangle \otimes \left| 
1 \right\rangle } \right)} 
\end{equation}

\begin{equation}
\label{ENheqn35}
Y\left| {W_6 } \right\rangle = \left| {W_7 } \right\rangle = a_3 
\sum\limits_{i = 0}^{N - 1} {''\left( {\left| i \right\rangle \otimes \left| 
0 \right\rangle } \right)} + b_3 \sum\limits_{i = 0}^{N - 1} {'\left( 
{\left| i \right\rangle \otimes \left| 0 \right\rangle } \right)} + c_3 
\sum\limits_{i = 0}^{N - 1} {'\left( {\left| i \right\rangle \otimes \left| 
1 \right\rangle } \right)} ,
\end{equation}

\noindent
where the mean used in the definition of partial diffusion operator is,

\begin{equation}
\label{ENheqn36}
\left\langle {\alpha _3 } \right\rangle = \frac{1}{N}\left( {\left( {N - M} 
\right)a_2 + Mc_2 } \right),
\end{equation}

\noindent
and $a_{3}$, $b_{3}$ and $c_{3}$ used in Eqn.(\ref{ENheqn35}) are calculated as follows:

\begin{equation}
\label{ENheqn37}
a_3 = 2\left\langle {\alpha _3 } \right\rangle - a_2 ;\,\,\,\,\,\,\,\,b_3 = 
2\left\langle {\alpha _3 } \right\rangle - c_2 ;\,\,\,\,\,\,c_3 = - b_2 ,
\end{equation}

\noindent
and the probabilities of the system are,

\begin{equation}
\label{ENheqn38}
\begin{array}{l}
 P_s^{(3)} = M\left( {b_3^2 + c_3^2 } \right) = M\left( {b_3^2 + b_2^2 } 
\right). \\ 
 P_{ns}^{(3)} = M\left( {a_3^2 } \right) = M\left( {b_3 + c_3 } \right)^2 = 
M\left( {b_3 - b_2 } \right)^2. \\ 
 \end{array}
\end{equation}

In general, the system after $q \ge 2$ iterations can be described using the 
following recurrence relations,

\begin{equation}
\label{ENheqn39}
\left| {W^{(q)}} \right\rangle = a_q \sum\limits_{i = 0}^{N - 1} {''\left( 
{\left| i \right\rangle \otimes \left| 0 \right\rangle } \right)} + b_q 
\sum\limits_{i = 0}^{N - 1} {'\left( {\left| i \right\rangle \otimes \left| 
0 \right\rangle } \right)} + c_q \sum\limits_{i = 0}^{N - 1} {'\left( 
{\left| i \right\rangle \otimes \left| 1 \right\rangle } \right)}, 
\end{equation}

\noindent
where the mean to be used in the definition of the partial diffusion 
operator is as follows: For simplicity, let $y = 1 - \frac{M}{N}$ and $s = 
\frac{1}{\sqrt N }$, then

\begin{equation}
\label{ENheqn40}
\left\langle {\alpha _q } \right\rangle = \left( {ya_{q - 1} + (1 - y)c_{q - 
1} } \right),
\end{equation}

\noindent
and $a_{q}$, $b_{q}$ and $c_{q}$ used in Eqn.(\ref{ENheqn39}) are calculated as follows:

\begin{equation}
\label{ENheqn41}
a_q = 2\left\langle {\alpha _q } \right\rangle - a_{q - 1} ;\,\,\,\,a_0 = 
s,\,\,\,\,a_1 = s\left( {2y - 1} \right),
\end{equation}

\begin{equation}
\label{ENheqn42}
b_q = 2\left\langle {\alpha _q } \right\rangle - c_{q - 1} ;\,\,\,\,\,b_0 = 
s,\,\,\,\,b_1 = 2sy,
\end{equation}

\begin{equation}
\label{ENheqn43}
c_q = - b_{q - 1} ;\,\,\,\,\,\,\,\,\,\,\,\,\,\,\,\,\,\,c_0 = 0,\,\,\,\,c_1 = 
- s,
\end{equation}

\noindent
and the probabilities of the system are,

\begin{equation}
\label{ENheqn44}
P_s^{(q)} = M\left( {b_q^2 + c_q^2 } \right) = M\left( {b_q^2 + b_{q - 1}^2 
} \right).
\end{equation}

\begin{equation}
\label{ENheqn45}
P_{ns}^{(q)} = M\left( {a_q^2 } \right) = M\left( {b_q + c_q } \right)^2 = 
M\left( {b_q - b_{q - 1} } \right)^2.
\end{equation}

For $q \ge 2$, Eqn.(\ref{ENheqn41}), Eqn.(\ref{ENheqn42}) and Eqn.(\ref{ENheqn43}) could 
be re-written as follows:

\begin{equation}
\label{ENheqn46}
a_q = 2ya_{q - 1} - a_{q - 2} ;\,\,\,\,\,a_0 = s;\,\,\,\,\,a_1 = s(2y - 1),
\end{equation}

\begin{equation}
\label{ENheqn47}
b_q = 2yb_{q - 1} - b_{q - 2} ;\,\,\,\,\,\,b_0 = s;\,\,\,\,\,b_1 = 2sy,
\end{equation}

\begin{equation}
\label{ENheqn48}
c_q = - b_{q - 1} ;\,\,\,\,\,\,\,\,\,\,\,\,\,\,\,\,\,\,\,\,\,c_0 = 
0;\,\,\,\,c_1 = - s,
\end{equation}

Solving the above recurrence relations (In Appendix A), 
the closed forms are as follows (Proved in Appendix B):

\begin{equation}
\label{ENheqn49}
a_q = s\left( {\frac{\sin \left( {\left( {q + 1} \right)\theta } 
\right)}{\sin \left( \theta \right)} - \frac{\sin \left( {q\theta } 
\right)}{\sin \left( \theta \right)}} \right),
\end{equation}

\begin{equation}
\label{ENheqn50}
b_q = s\left( {\frac{\sin \left( {\left( {q + 1} \right)\theta } 
\right)}{\sin \left( \theta \right)}} \right),
\end{equation}

\begin{equation}
\label{ENheqn51}
c_q = - s\left( {\frac{\sin \left( {q\theta } \right)}{\sin \left( \theta 
\right)}} \right),
\end{equation}

\noindent
where $y = \cos \left( \theta \right)$ and $0 < \theta \le \frac{\pi }{2}$. 
The above closed forms can be expressed via the Chebyshev polynomials of the 
second kind $U_q \left( y \right)$ \cite{ChebPoly}, which 
are defined as follows, 

\begin{equation}
\label{ENheqn52}
U_q \left( y \right) = \frac{\sin \left( {\left( {q + 1} \right)\theta } 
\right)}{\sin \left( \theta \right)}.
\end{equation}

This allows us to re-write the above closed form in terms of Chebyshev 
polynomials of the second kind as follows,

\begin{equation}
\label{ENheqn53}
a_q = s\left( {U_q - U_{q - 1} } \right),
\end{equation}

\begin{equation}
\label{ENheqn54}
b_q = sU_q, 
\end{equation}

\begin{equation}
\label{ENheqn55}
c_q = - sU_{q - 1}, 
\end{equation}

And the probabilities of the system,

\begin{equation}
\label{ENheqn56}
P_s^{(q)} = (1 - \cos \left( \theta \right))\left( {U_q^2 + U_{q - 1}^2 } 
\right).
\end{equation}

\begin{equation}
\label{ENheqn57}
P_{ns}^{(q)} = \cos \left( \theta \right)\left( {U_q - U_{q - 1} } 
\right)^2.
\end{equation}

Such that,

\begin{equation}
\label{ENheqn58}
\begin{array}{l}
 P_s^{(q)} + P_{ns}^{(q)} = M\left( {b_q^2 + c_q^2 } \right) + \left( {N - 
M} \right)a_q^2 \\ 
\,\,\,\,\,\,\,\,\,\,\,\,\,\, = N\left( {b_q^2 + c_q^2 } 
\right) + 2\left( {N - M} \right)c_q b_q \\ 
\,\,\,\,\,\,\,\,\,\,\,\,\,\, = \frac{1}{\sin ^2\left( \theta 
\right)}\left( {\sin ^2\left( {\left( {j + 1} \right)\theta } \right) + \sin 
^2\left( {j\theta } \right) - 2\cos \left( \theta \right)\sin \left( {\left( 
{j + 1} \right)\theta } \right)\sin \left( {j\theta } \right)} \right) \\ 
 \,\,\,\,\,\,\,\,\,\,\,\,\,\, = \frac{1}{\sin ^2\left( \theta 
\right)}\left( {\cos ^2\left( {j\theta } \right)\sin ^2\left( \theta \right) 
- \sin ^2\left( {j\theta } \right)\cos ^2\left( \theta \right) + \sin 
^2\left( {j\theta } \right)} \right) \\ 
 \,\,\,\,\,\,\,\,\,\,\,\,\,\, = \frac{1}{\sin ^2\left( \theta 
\right)}\left( {\left( {1 - \sin ^2\left( {j\theta } \right)} \right)\sin 
^2\left( \theta \right) - \sin ^2\left( {j\theta } \right)\left( {1 - \sin 
^2\left( \theta \right)} \right) + \sin ^2\left( {j\theta } \right)} \right) 
\\ 
 \,\,\,\,\,\,\,\,\,\,\,\,\,\, = \frac{\sin ^2\left( \theta 
\right)}{\sin ^2\left( \theta \right)} = 1. \\ 
 \end{array}
\end{equation}

\subsubsection{Performance of Iterating the Algorithm}

Now, we have to calculate how many iterations, $q$, are required to find the 
matches with certainty or near certainty for different cases of $1 \le M 
\le N$. To find a match with certainty on any measurement, then $P_s^{(q)} $ must be 
as close as possible to certainty. To calculate the number of iterations, 
$q$, required to satisfy this condition, we need the following theorem.

\begin{theorem} 

Consider the following relation,

\begin{equation}
\label{ENheqn59}
(1 - \cos \left( \theta \right))\left( {U_q^2 + U_{q - 1}^2 } \right) = 1,
\end{equation}

\noindent
where $U_q \left( y \right)$ is Chebyshev polynomials of the second kind, $y 
= \cos \left( \theta \right)$ and $0 < \theta \le \frac{\pi }{2}$, then, 

\[
q = \frac{\pi - \theta }{2\theta } \mbox{ or } \theta = \frac{\pi }{2}.
\]

\begin{proof}

From the definition of $U_q $ shown in Eqn.(\ref{ENheqn52}) then Eqn.(\ref{ENheqn59}) can take this 
form,

\[
\left( {1 - \cos \left( \theta \right)} \right)\left( {\frac{\sin ^2\left( 
{\left( {q + 1} \right)\theta } \right)}{\sin ^2\left( \theta \right)} + 
\frac{\sin ^2\left( {q\theta } \right)}{\sin ^2\left( \theta \right)}} 
\right) = 1,
\]

\noindent
or,

\[
\sin ^2\left( {\left( {q + 1} \right)\theta } \right) + \sin ^2\left( 
{q\theta } \right) = 1 + \cos \left( \theta \right).
\]

Using simple trigonometric identities, the above relation may take the form,

\[
\cos \left( {2q\theta + 2\theta } \right) + \cos \left( {2q\theta } \right) 
+ 2\cos \left( \theta \right) = 0.
\]

Using the addition formulas for cosine we get,

\[
2\cos \left( {2q\theta } \right)\cos ^2\left( \theta \right) - 2\cos \left( 
\theta \right)\sin \left( {2q\theta } \right)\sin \left( \theta \right) + 
2\cos \left( \theta \right) = 0,
\]

\[
2\cos \left( \theta \right)\left( {\cos \left( {2q\theta } \right)\cos 
\left( \theta \right) - \sin \left( {2q\theta } \right)\sin \left( \theta 
\right) - 1} \right) = 0,
\]

\[
\cos \left( \theta \right)\left( {\cos \left( {2q\theta + \theta } \right) - 
1} \right) = 0.
\]

From the last equation we get,
\[
\cos \left( \theta \right) = 0 \mbox{ or } \cos \left( {2q\theta + \theta } \right) = 
\cos \left( { - \pi } \right),
\]

\noindent
which gives the required conditions,

\[
\theta = \frac{\pi }{2} \mbox{ or } q = \frac{\pi - \theta }{2\theta }.
\]
\end{proof}
\end{theorem}

Using the above result, and since the number of iterations must be integer, 
following the same fashion as shown in \cite{boyer96} , 
then the required number of iterations is,

\begin{equation}
\label{ENheqn60}
q = \left\lfloor {\frac{\pi }{2\sqrt 2 }\sqrt {\frac{N}{M}} } \right\rfloor, 
\end{equation}

\noindent
where $\left\lfloor {\,\,\,} \right\rfloor $ is the floor operation. The 
algorithm runs in $O\left( {\sqrt {N / M} } \right)$ with no contradiction 
with the prove of optimality shown in \cite{boyer96,niel00}.

\section{Comparison with Grover's Algorithm}

First we will summarize the above results from both Grover's and the 
proposed algorithm before starting the comparison. The probability of 
success of Grover's algorithm as shown in \cite{boyer96} is as follows:

\begin{equation}
\label{ENheqn61}
P_s^{\left( {q_G } \right)} = \sin ^2\left( {\left( {2q_G + 1} \right)\theta 
} \right),
\end{equation}

\noindent
where $\sin ^2\left( \theta \right) = \frac{M}{N}\,\,\,\,;0 < 
\theta \le \frac{\pi }{2}$ and the required $q_G$ is,

\begin{equation}
\label{ENheqn62}
q_G = \left\lfloor {\frac{\pi }{4}\sqrt {\frac{N}{M}} } \right\rfloor. 
\end{equation}

For the proposed algorithm, the probability of success is as follows,

\begin{equation}
\label{ENheqn63}
P_s^{(q)} = \left( {1 - \cos \left( \theta \right)} \right)\left( 
{\frac{\sin ^2\left( {\left( {q + 1} \right)\theta } \right)}{\sin ^2\left( 
\theta \right)} + \frac{\sin ^2\left( {q\theta } \right)}{\sin ^2\left( 
\theta \right)}} \right),
\end{equation}

\noindent
where $\cos \left( \theta \right) = 1-\frac{M}{N};0 < \theta \le \frac{\pi }{2}$, and 
the required $q$ is,

\begin{equation}
\label{ENheqn64}
q = \left\lfloor {\frac{\pi }{2\sqrt 2 }\sqrt {\frac{N}{M}} } \right\rfloor. 
\end{equation}

\begin{figure}[htbp]
\centerline{\includegraphics[width=4.167in,height=7.139in]{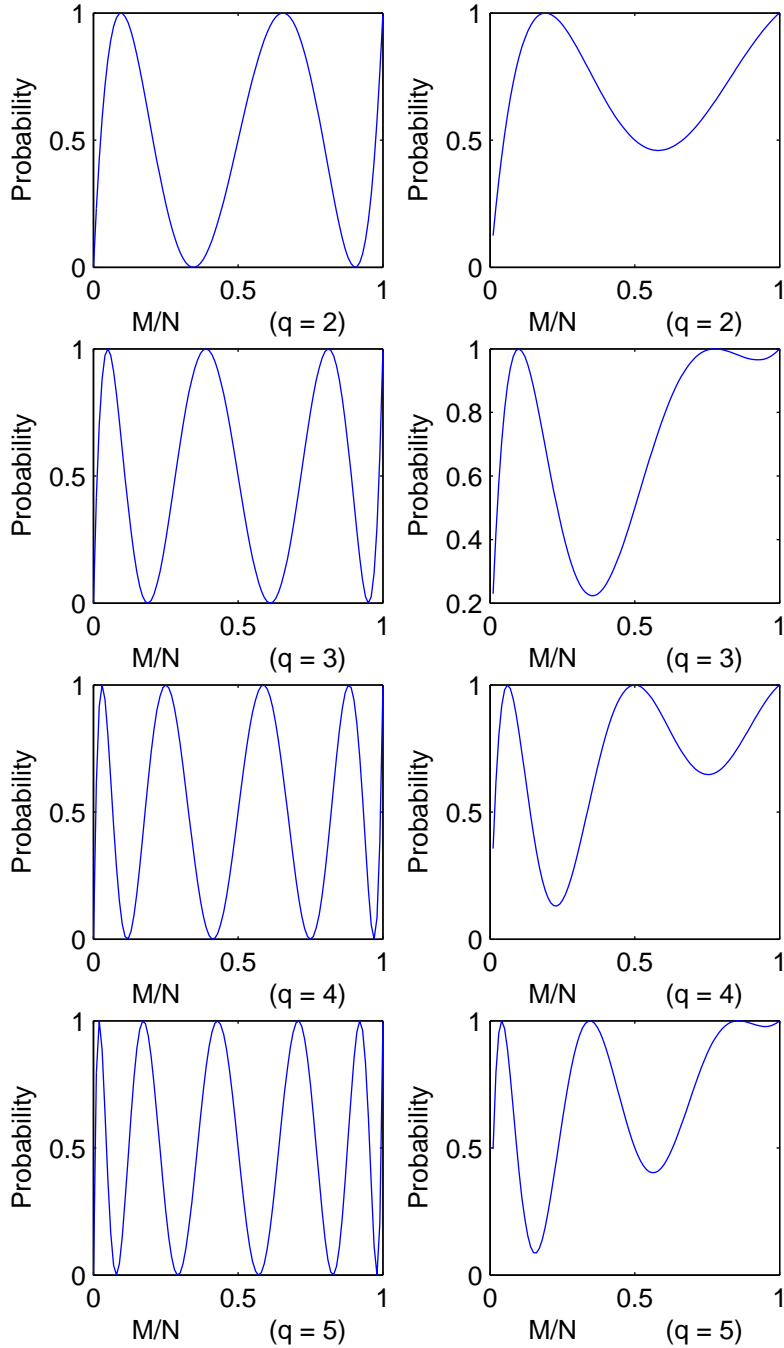}}
\caption{Probability of success as a function of $0<\frac{M}{N}\le1$ after different number of iterations: Grover's algorithm (left) vs. the proposed algorithm (right).} 
\label{comp25}
\end{figure}

Fig.(\ref{comp25}) shows the probability of success as a function of the ratio $M/N$ for both 
algorithms, after 2, 3, 4 and 5 iterations. It is clear from the graphs of the proposed algorithm that 
the probability will never return to {\it zero} once started and the minimum probability will increase as 
$M$ increases because of the use of the partial diffusion operator which will resist the de-amplification when reaching 
the turning points as explained in the definition of the partial diffusion operator, i.e. the problem becomes easier for 
multiple matches, where for Grover's algorithm, the number of cases (points) to be solved with certainty is equal to the number of 
cases with zero-probability after arbitrary number of iterations.

\begin{figure}[htbp]
\centerline{\includegraphics[width=3.622in,height=2.8171in]{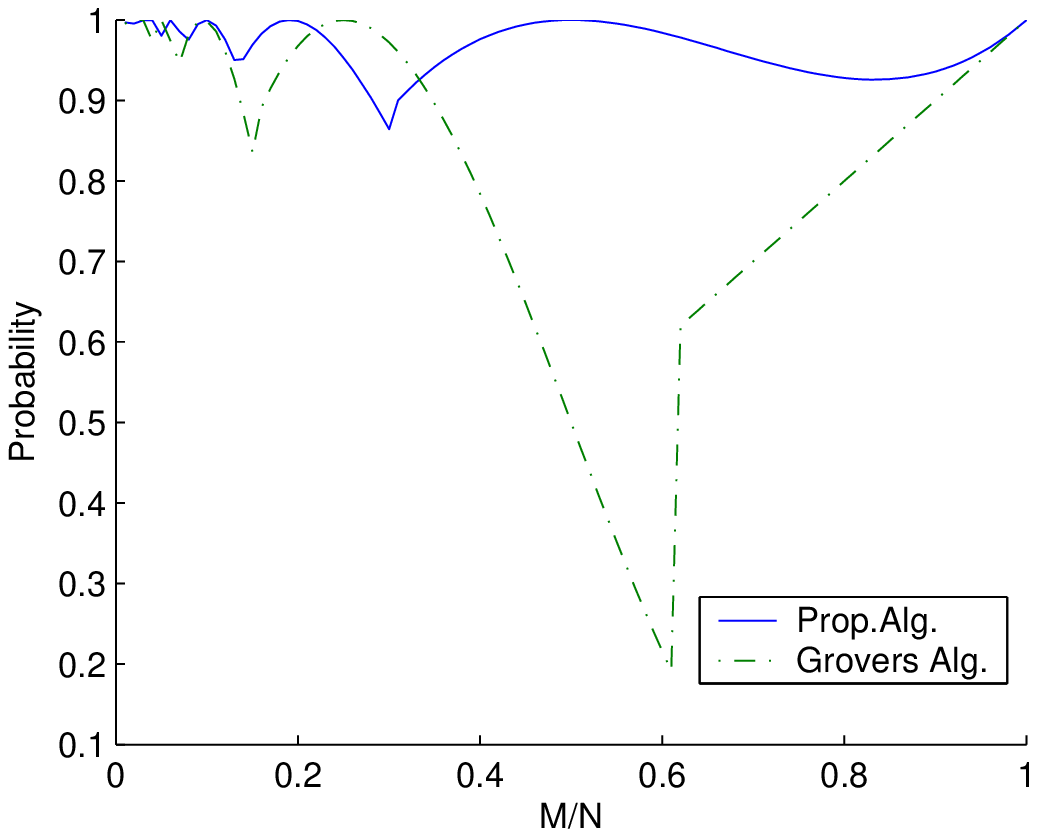}}
\caption{Probability distribution using the appropriate number 
of iterations for both algorithms.} 
\label{jitr}
\end{figure}

\begin{figure}[htbp]
\centerline{\includegraphics[width=3.567in,height=2.775in]{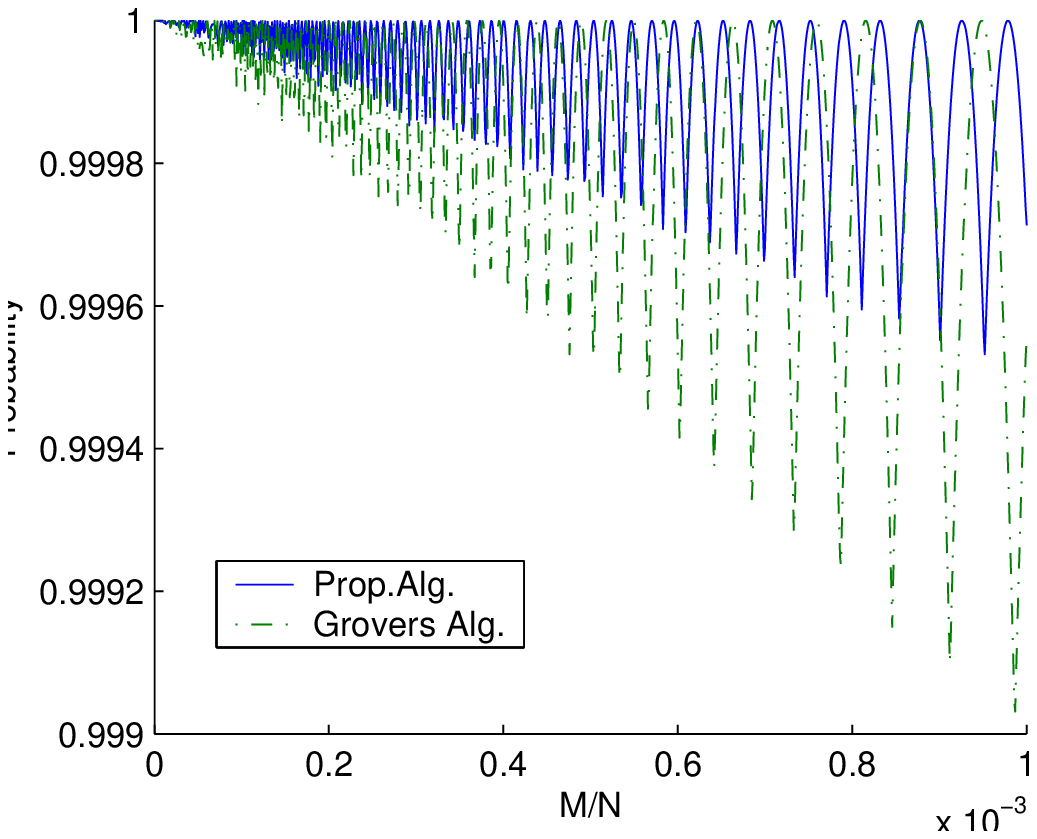}}
\caption{Probability distribution using the appropriate number 
of iterations for both algorithms for the hardest cases where$M / N < 
1\times 10^{ - 3}$.} 
\label{JItrsmall}
\end{figure}

Another way to understand the behaviour of both algorithms is to plot the 
probability of success using the calculated number of iterations for each 
algorithm, Fig.(\ref{jitr}) shows the probability of success as a function of the 
ratio $M/N$ for both algorithms by inserting the calculated number of iterations 
$q_G $ and $q$ shown in Eqn.(\ref{ENheqn62}) and Eqn.(\ref{ENheqn64}) in $P_s^{(q)} $ and $P_s^{(q_G )} 
$ respectively. We can see from the plot that the minimum probability that 
Grover's algorithm may reach is approx. 17.49 {\%} when $M/N= 0.61685$ while for 
the proposed algorithm, the minimum probability is 84.72{\%} when $M/N = 
0.30842$. Fig.(\ref{JItrsmall}) shows the same behaviour for both 
algorithms for small $M$ and large $N$ (hard cases where $M / N < 1\times 10^{ -3})$. 
It is interesting to notice that the behaviour for the proposed 
algorithm shown in Fig.(\ref{jitr}) is similar to the behaviour of the first iteration 
shown in Fig.(\ref{ENhplot}) for $M/N > 0.30842$ which implies that if $M/N > 0.30842$ then 
the proposed algorithm runs in $O(1)$, i.e. the problem is easier for 
multiple matches.

\section{Conclusion}

In this paper, we presented a quantum algorithm for searching unstructured 
list of size $N$ runs in $O\left( {\sqrt {N / M} } \right)$ where $M$ is the number of 
matches within the list. The algorithm operations based on the {\it Partial Diffusion Operator} works similar to 
the Diffusion Operator used in Grover's algorithm \cite{grover96} except that it 
performs the inversion about the mean 
only {\it on a subspace of the system}. Using this 
operator, we showed that the algorithm performs {\it more 
reliable} than Grover's algorithm in case of fewer number of matches 
(hard cases of the problem) and runs in $O(1)$ in case of multiple matches 
(easy cases of the problem).

\section*{Appendix A}

In this appendix, we will solve the following recurrence relations shown in 
Eqn.(\ref{ENheqn46}) and Eqn.(\ref{ENheqn47}) to get their close forms: First for $a_q$ which is defined as follows 
for $q\ge2$,
\begin{equation}
\label{Enhappa1}
a_q = 2ya_{q - 1} - a_{q - 2} ;\,\,q \ge 2 \mbox{ such that }, a_0 = s;\,\,a_1 = s(2y - 1).
\end{equation}

The characteristic equation,

\begin{equation}
\label{Enhappa2}
\lambda ^2 - 2y\lambda + 1 = 0,
\end{equation}
\noindent
and,

\begin{equation}
\label{Enhappa3}
\lambda _{1,2} = y\pm i\sqrt {1 - y^2}. 
\end{equation}

Let $y = \cos \left( \theta \right)$ such that $0 < \theta \le \frac{\pi 
}{2}$, then,

\begin{equation}
\label{Enhappa4}
\lambda _{1,2} = \cos \left( \theta \right)\pm i\sin \left( \theta \right) = 
e^{\pm i\theta }.
\end{equation}

So, the closed form will take the form,

\begin{equation}
\label{Enhappa5}
a_q = \upsilon _1 e^{iq\theta } + \upsilon _2 e^{ - iq\theta },
\end{equation}

\noindent
where $\upsilon _1 $ and $\upsilon _2 $ are constants to be determined from 
the initial conditions as follows,

\begin{equation}
\label{Enhappa6}
\upsilon _1 = \frac{s\left( {e^{ - i\theta } - 2y + 1} \right)}{e^{ - 
i\theta } - e^{i\theta }};\,\,\,\,\,\,\upsilon _2 = \frac{s\left( {2y - 1 - 
e^{i\theta }} \right)}{e^{ - i\theta } - e^{i\theta }}.
\end{equation}

Substituting in Eqn.(\ref{Enhappa5}) we get,

\begin{equation}
\label{Enhappa7}
a_q = \frac{s}{\sin \left( \theta \right)}\left( {2\cos \left( \theta 
\right)\sin \left( {q\theta } \right) - \sin \left( {\left( {q - 1} 
\right)\theta } \right) - \sin \left( {q\theta } \right)} \right).
\end{equation}

We have from the identities of multiple-angle formulas,

\begin{equation}
\label{Enhappa8}
\sin \left( {\left( {q + 1} \right)\theta } \right) = 2\cos \left( \theta 
\right)\sin \left( {q\theta } \right) - \sin \left( {\left( {q - 1} 
\right)\theta } \right).
\end{equation}

So the closed form of $a_q $ may take the following form,

\begin{equation}
\label{Enhappa9}
a_q = s\left( {\frac{\sin \left( {\left( {q + 1} \right)\theta } 
\right)}{\sin \left( \theta \right)} - \frac{\sin \left( {q\theta } 
\right)}{\sin \left( \theta \right)}} \right).
\end{equation}

Second, for $b_q$ which is defined as follows for $q\ge2$,

\begin{equation}
\label{Enhappa10}
b_q = 2yb_{q - 1} - b_{q - 2}; \mbox{ such that, } b_0 = s;\,\,b_1 = 2sy.
\end{equation}

Since we are starting with the same recurrence relation used for $a_q$ but with 
different initial conditions, then the closed form for $b_q $ may take the 
form,

\begin{equation}
\label{Enhappa11}
b_q = \upsilon _1 e^{iq\theta } + \upsilon _2 e^{ - iq\theta },
\end{equation}

\noindent
where $\upsilon _1 $ and $\upsilon _2 $ are constants to be determined from 
the initial conditions as follows,

\begin{equation}
\label{Enhappa12}
\upsilon _1 = \frac{se^{ - i\theta } - 2ys}{e^{ - i\theta } - e^{i\theta 
}};\,\,\,\,\,\,\upsilon _2 = \frac{2ys - se^{i\theta }}{e^{ - i\theta } - 
e^{i\theta }}.
\end{equation}

Substituting in Eqn.(\ref{Enhappa11}) we get,

\begin{equation}
\label{Enhappa13}
b_q = \frac{s}{\sin \left( \theta \right)}\left( {2\cos \left( \theta 
\right)\sin \left( {q\theta } \right) - \sin \left( {\left( {q - 1} 
\right)\theta } \right)} \right).
\end{equation}

So the closed form of $b_q $ may take the following form,

\begin{equation}
\label{Enhappa14}
b_q = s\left( {\frac{\sin \left( {\left( {q + 1} \right)\theta } 
\right)}{\sin \left( \theta \right)}} \right).
\end{equation}

\section*{Appendix B}

In this appendix, we want to prove that, for $q \ge 2$, if the amplitudes $a_q $, $b_q$ and 
$c_q $ used in Eqn.(\ref{ENheqn39}) are defined according to the definition of Partial 
diffusion operator shown in Eqn.(\ref{ENheq13}) as follows,

\begin{equation}
\label{Enhappb1}
\left\langle {\alpha _q } \right\rangle = \left( {ya_{q - 1} + (1 - y)c_{q - 
1} } \right),
\end{equation}

\begin{equation}
\label{Enhappb2}
a_q = 2\left\langle {\alpha _q } \right\rangle - a_{q - 1} ;\,\,\,\,a_0 = 
s,\,\,\,\,a_1 = s\left( {2y - 1} \right),
\end{equation}

\begin{equation}
\label{Enhappb3}
b_q = 2\left\langle {\alpha _q } \right\rangle - c_{q - 1} ;\,\,\,\,\,b_0 = 
s,\,\,\,\,b_1 = 2sy,
\end{equation}

\begin{equation}
\label{Enhappb4}
c_q = - b_{q - 1} ;\,\,\,\,\,\,\,\,\,\,\,\,\,\,\,\,\,\,c_0 = 0,\,\,\,\,c_1 = 
- s,
\end{equation}

\noindent
then their closed forms are as follows, where $y = \cos \left( \theta 
\right)$ and $0 < \theta \le \frac{\pi }{2}$:

\begin{equation}
\label{Enhappb5}
a_q = s\left( {\frac{\sin \left( {\left( {q + 1} \right)\theta } 
\right)}{\sin \left( \theta \right)} - \frac{\sin \left( {q\theta } 
\right)}{\sin \left( \theta \right)}} \right),
\end{equation}

\begin{equation}
\label{Enhappb6}
b_q = s\left( {\frac{\sin \left( {\left( {q + 1} \right)\theta } 
\right)}{\sin \left( \theta \right)}} \right),
\end{equation}

\begin{equation}
\label{Enhappb7}
c_q = - s\left( {\frac{\sin \left( {q\theta } \right)}{\sin \left( \theta 
\right)}} \right).
\end{equation}

\begin{proof}

Substituting $y = \cos \left( \theta \right)$ in the definition and eliminating $c_q$ 
since it is sufficient to prove the closed forms for $a_q$ and $b_q$ we get,

\[
a_q = \left( {2\cos \left( \theta \right) - 1} \right)a_{q - 1} - 2\left( {1 
- \cos \left( \theta \right)} \right)b_{q - 2} ,
\]

\[
b_q = 2\cos \left( \theta \right)a_{q - 1} - \left( {1 - 2\cos \left( \theta 
\right)} \right)b_{q - 2} ,
\]

\noindent
with initial conditions,

\[
a_0 = s;\,\,a_1 = s(2\cos \left( \theta \right) - 1),
\]

\[
b_0 = s;\,\,\,\,\,b_1 = 2s\cos \left( \theta \right),
\]

{\bf Step 1:} Prove for $q=2$. 

For $a_{2}$, from definition and initial conditions,

\[
\begin{array}{l}
 a_2 = \left( {2\cos \left( \theta \right) - 1} \right)a_1 - 2\left( {1 - 
\cos \left( \theta \right)} \right)b_0 \\ 
 \,\,\,\,\,\, = \left( {2\cos \left( \theta \right) - 1} \right)\left( 
{2\cos \left( \theta \right) - 1} \right)s - 2\left( {1 - \cos \left( \theta 
\right)} \right)s \\ 
 \,\,\,\,\,\, = s\left( {\left( {2\cos \left( \theta \right) - 1} \right)^2 
- 2\left( {1 - \cos \left( \theta \right)} \right)} \right) \\ 
 \,\,\,\,\,\, = s\left( {4\cos ^2\left( \theta \right) - 2\cos \left( \theta 
\right) - 1} \right) \\ 
 \,\,\,\,\,\, = s\left( {3\cos ^2\left( \theta \right) - \sin ^2\left( 
\theta \right) - 2\cos \left( \theta \right)} \right) \\ 
 \,\,\,\,\,\, = s\left( {\frac{3\cos ^2\left( \theta \right)\sin \left( 
\theta \right) - \sin ^3\left( \theta \right) - 2\cos \left( \theta 
\right)\sin \left( \theta \right)}{\sin \left( \theta \right)}} \right) \\ 
 \,\,\,\,\,\, = s\left( {\frac{\sin \left( {3\theta } \right)}{\sin \left( 
\theta \right)} - \frac{\sin \left( {2\theta } \right)}{\sin \left( \theta 
\right)}} \right). \\ 
 \end{array}
\]

For $b_{2}$, from definition and initial conditions,

\[
\begin{array}{l}
 b_2 = 2\cos \left( \theta \right)\left( {2\cos \left( \theta \right) - 1} 
\right)s - \left( {1 - 2\cos \left( \theta \right)} \right)s \\ 
 \,\,\,\,\, = s\left( {4\cos ^2\left( \theta \right) - 1} \right) \\ 
 \,\,\,\,\, = s\left( {3\cos ^2\left( \theta \right) - \sin ^2\left( \theta 
\right)} \right) \\ 
 \,\,\,\,\, = s\left( {\frac{3\cos ^2\left( \theta \right)\sin \left( \theta 
\right) - \sin ^3\left( \theta \right)}{\sin \left( \theta \right)}} \right) 
\\ 
 \,\,\,\, = s\left( {\frac{\sin \left( {3\theta } \right)}{\sin \left( 
\theta \right)}} \right). \\ 
 \end{array}
\]

{\bf Step 2}: Assume the relation is true for $q=t-1$ and $q=t$,

\[
a_{t - 1} = s\left( {\frac{\sin \left( {t\theta } \right)}{\sin \left( 
\theta \right)} - \frac{\sin \left( {\left( {t - 1} \right)\theta } 
\right)}{\sin \left( \theta \right)}} \right);\,\,\,\,a_t = s\left( 
{\frac{\sin \left( {\left( {t + 1} \right)\theta } \right)}{\sin \left( 
\theta \right)} - \frac{\sin \left( {t\theta } \right)}{\sin \left( \theta 
\right)}} \right).
\]

\[
b_{t - 1} = s\left( {\frac{\sin \left( {t\theta } \right)}{\sin \left( 
\theta \right)}} \right);b_t = s\left( {\frac{\sin \left( {\left( {t + 1} 
\right)\theta } \right)}{\sin \left( \theta \right)}} \right).
\]

{\bf Step 3}: Prove for $q=t+1$,

For $a_{t + 1}$, from the definition and assumption,

\[
\begin{array}{l}
 a_{t + 1} = \left( {2\cos \left( \theta \right) - 1} \right)a_t - 2\left( 
{1 - \cos \left( \theta \right)} \right)b_{t - 1} \\ 
 \,\,\,\,\,\,\,\,\, = \left( {2\cos \left( \theta \right) - 1} 
\right)s\left( {\frac{\sin \left( {\left( {t + 1} \right)\theta } 
\right)}{\sin \left( \theta \right)} - \frac{\sin \left( {t\theta } 
\right)}{\sin \left( \theta \right)}} \right) - 2\left( {1 - \cos \left( 
\theta \right)} \right)s\left( {\frac{\sin \left( {t\theta } \right)}{\sin 
\left( \theta \right)}} \right) \\ 
 \,\,\,\,\,\,\,\,\, = \frac{s}{\sin \left( \theta \right)}\left( {2\cos 
\left( \theta \right)\sin \left( {\left( {t + 1} \right)\theta } \right) - 
\sin \left( {t\theta } \right) - \sin \left( {\left( {t + 1} \right)\theta } 
\right)} \right) \\ 
 \,\,\,\,\,\,\,\,\, = s\left( {\frac{\sin \left( {\left( {t + 2} 
\right)\theta } \right)}{\sin \left( \theta \right)} - \frac{\sin \left( 
{\left( {t + 1} \right)\theta } \right)}{\sin \left( \theta \right)}} 
\right). \\ 
 \end{array}
\]

For $b_{t + 1}$, from the definition and assumption,

\[
\begin{array}{l}
 b_{t + 1} = 2\cos \left( \theta \right)a_t - \left( {1 - 2\cos \left( 
\theta \right)} \right)b_{t - 1} \\ 
 \,\,\,\,\,\,\,\, = 2\cos \left( \theta \right)s\left( {\frac{\sin \left( 
{\left( {t + 1} \right)\theta } \right)}{\sin \left( \theta \right)} - 
\frac{\sin \left( {t\theta } \right)}{\sin \left( \theta \right)}} \right) - 
\left( {1 - 2\cos \left( \theta \right)} \right)s\left( {\frac{\sin \left( 
{t\theta } \right)}{\sin \left( \theta \right)}} \right) \\ 
 \,\,\,\,\,\,\,\, = \frac{s}{\sin \left( \theta \right)}\left( {2\cos \left( 
\theta \right)\sin \left( {\left( {t + 1} \right)\theta } \right) - \sin 
\left( {t\theta } \right)} \right) \\ 
 \,\,\,\,\,\,\,\,\, = s\left( {\frac{\sin \left( {\left( {t + 2} 
\right)\theta } \right)}{\sin \left( \theta \right)}} \right). \\ 
 \end{array}
\]

\noindent
and this completes the proof.

\end{proof}

\end{document}